\documentclass[useAMS,usenatbib]{mn2e}
\usepackage{txfonts}
\usepackage{natbib}
\usepackage{aas_macros}
\usepackage{graphicx}
\usepackage{subfigure}

\newcommand{\gsim}{\mathrel{\hbox{\rlap{\lower.55ex \hbox {$\sim$}}
                   \kern-.3em \raise.4ex \hbox{$>$}}}}
\newcommand{\lsim}{\mathrel{\hbox{\rlap{\lower.55ex \hbox {$\sim$}}
                   \kern-.3em \raise.4ex \hbox{$<$}}}}

\title[Circumplanetary disc properties]{Circumplanetary disc properties obtained from radiation hydrodynamical simulations of gas accretion by protoplanets}
\author[B.A. Ayliffe \& M.R. Bate]{Ben A. Ayliffe\thanks{E-mail:
ayliffe@astro.ex.ac.uk} and Matthew R. Bate\thanks{E-mail:
mbate@astro.ex.ac.uk}\\ School of Physics, University of Exeter, Stocker
Road, Exeter EX4 4QL}

\date{\today}
\begin{document}
\maketitle
\begin{abstract}

We investigate the properties of circumplanetary discs formed in three-dimensional, self-gravitating radiation hydrodynamical models of gas accretion by protoplanets. We determine disc sizes, scaleheights, and density and temperature profiles for different protoplanet masses, in solar nebulae of differing grain opacities.

We find that the analytical prediction of circumplanetary disc radii in an evacuated gap ($R_{\rm Hill}/3$) from \cite{QuilTril1998} yields a good estimate for discs formed by high mass protoplanets. The radial density profiles of the circumplanetary discs may be described by power-laws between $r^{-2}$ and $r^{-3/2}$. We find no evidence for the ring-like density enhancements that have been found in some previous models of circumplanetary discs. Temperature profiles follow a $\sim r^{-7/10}$ power-law regardless of protoplanet mass or nebula grain opacity. The discs invariably have large scaleheights ($H/r > 0.2$), making them thick in comparison with their encompassing circumstellar discs, and they show no flaring.

\end{abstract}
\begin{keywords}
planets and satellites: formation -- accretion, accretion discs -- hydrodynamics -- radiative transfer -- (stars:) planetary systems: formation -- methods: numerical
\end{keywords}

\section{Introduction}

The giant planets of our solar system all have multiple satellites, divided into two classes; the regular and irregular satellites. The regular satellites have low eccentricity orbits with low inclinations relative to the planet's equatorial plane \citep{Galileo1610, Galileo, MosEst2003}, and possess prograde orbits. These characteristics are suggestive of satellite formation in a disc of material rotating about the planet. Contrastingly, the irregular satellites have eccentric, highly inclined orbits, and show no strong bias between prograde or retrograde orbits \citep{GraHolGlaAks2003}. These characteristics suggest that the irregular satellites are captured by their host planets, rather than formed in situ. However, this capture process requires a mechanism by which to lose energy, for example gas drag as suggested by \cite*{PolBurTau1979}. This was more recently revisited by \cite{CukBur2004}, who modify the original scenario to include gas drag from a circumplanetary disc as opposed to the spherical envelope assumed by \citeauthor{PolBurTau1979}. \citeauthor{CukBur2004} find that drag due to the circumplanetary disc gas is sufficient to enable the capture of bodies up to 100km in radius, which would otherwise form only temporary associations. Therefore circumplanetary discs may be instrumental in the formation of satellite systems comprising both regular and irregular satellites.

To form regular satellites a disc must exist around a protoplanet in which agglomeration processes can occur. \cite{CanWar2002} discuss the uncertainty in the nature of such a disc, whether it is an accretion circumplanetary disc or a spin-out disc. The former is created when a protoplanet begins to contract whilst still embedded in the circumstellar disc. When the protoplanet attains a sufficiently small radius it becomes impossible for the high angular momentum gas from the circumstellar disc to fall directly onto its surface. Instead it forms an accretion disc through which angular momentum can be dispersed before the gas infalls to the protoplanet's atmosphere. A spin-out disc might form when a protoplanet contracts after the circumstellar disc has dispersed. In this scenario, as the protoplanet contracts, (as there is no ambient environment into which the angular momentum may be disseminated) so its rotation rate increases. Upon becoming rotationally unstable, the most rapidly rotating material which occurs at equatorial latitudes, is flung out to form a spin-out disc.

One-dimensional models suggest that a Jupiter mass protoplanet will contract to less than 1/100th of its Hill radius in under 1Myr \citep{papa2005}. This would be within the circumstellar disc lifetime, allowing the protoplanet to continue to accrete material via an accretion disc. This suggests that it is an accretion disc in which satellites form rather than in a spin-out disc. Indeed, previous two-dimensional \citep*{LubSeiArt1999,DAnHenKley2002} and three-dimensional (\citealt{batezeus}; \citealt*{angelo2003}; \citealt{FouMay2008,AylBate2009}) hydrodynamical models of planet formation have shown the existence of circumplanetary accretion discs during giant planet formation. The two-dimensional models that develop circumplanetary discs have shown strong shocks within these discs \citep{LubSeiArt1999,DAnHenKley2002} that rapidly deplete them of material, which accretes on to the protoplanet. \cite{batezeus} showed that in three-dimensional models these shocks appear to be much weaker than in two dimensions, suggesting a slower rate of circumplanetary disc depletion. Such weakened shocks also appear in recent three-dimensional calculations by \cite{Machida2008}. Following dispersal of the circumstellar disc, the circumplanetary disc is no longer replenished with gas and solids. The residual circumplanetary disc must presumably survive for a considerable period to allow satellites to form from the volatile compounds that will only solidify at low temperatures which cannot be achieved in an actively accreting disc. A long disc lifetime also allows time for the constituent materials (ice, rock, etc.) to agglomerate, and to do so slowly enough that the collisional heating does not sublimate the ice \citep{CanWar2008}.

In this paper, we present and discuss the properties of circumplanetary discs formed in three-dimensional radiation hydrodynamical calculations of giant planet growth. These models supersede the previous three-dimensional models mentioned above in that they include radiative transfer. Therefore, they provide a more accurate description of the early phase of circumplanetary discs, in which satellites are believed to form. We examine the radial extents, density profiles, temperatures, and scaleheights of discs around different mass protoplanets, in nebulae with different grain opacities. In Section 2, we describe our computational method. Section 3 presents the results from our self-gravitating radiation hydrodynamical simulations, while in Section 4 we discuss our results in the context of previous work. Our conclusions are given in Section 5.

\section{Computational method}
\label{method}

The calculations described herein have been performed using a three-dimensional smoothed particle hydrodynamics (SPH) code. This SPH code has its origins in a version first developed by \citeauthor{benz90} (\citeyear{benz90}; \citealt{benzcam90}) but it has undergone substantial modification in the subsequent years. Energy and entropy are conserved to timestepping accuracy by use of the variable smoothing length formalism of \cite{springel2002} and \cite{Monaghan2002} with our specific implementation being described in \cite{price2007}. Gravitational forces are calculated and neighbouring particles are found using a binary tree. Radiative transfer is modelled in the flux-limited diffusion approximation using the method developed by \citet*{WhiBatMon2005} and \citet{WhiBat2006}.  Integration of the SPH equations is achieved using a second-order Runge-Kutta-Fehlberg integrator with particles having individual timesteps \citep*{bate95}. The standard implementation of artificial viscosity is used \citep{mongin83,mon92} with the parameters $\alpha_{v}$ = 1 and $\beta_{v}$ = 2. The code has been parallelised by M. Bate using OpenMP.

The calculations used in this paper have been performed in the same way, and in some instances are the same as calculations performed in \cite{AylBate2009} (henceforth Paper I). It is therefore only in brief that the details of the method are outlined below, with more details available in Paper I.

\subsection{Model setup}
\label{sec:model}

Our calculations are performed in the reference frame of the planet, orbiting a star of mass $M_{*}$ at radius $r_{\rm p}$ with an angular speed given by $\Omega_{\rm p} = \sqrt{GM_{*}/r_{p}^{3}}$, neglecting the mass of the planet.  The stellar mass is taken as 1$~\rm M_{\sun}$ and the orbital radius of the planet to be 5.2 AU.  We model only a small section of the protoplanetary disc centred on the protoplanet. Our standard section size is $r=1\pm 0.15~r_{\rm p}$ ($5.2 \pm 0.78~{\rm AU}$), and $\phi = \pm 0.15$ radians.  The initial radial temperature profile for the disc is $T_{\rm g} \propto r^{-1}$, which leads to a constant ratio of disc scaleheight with radius of $H/r=0.05$. The initial surface density of the disc has as $\Sigma \propto r^{-1/2}$ profile, with a value of 75~$\rm g~cm^{-2}$ at the planet's orbital radius.

\subsection{Boundary conditions}

The distribution of particles within the disc section is initially that of an unperturbed disc, with Keplerian velocities. Inward and outward of the planet's orbital radius the particles flow in opposite directions in the reference frame of the planet, and therefore leave the modelled section by opposite azimuthal boundaries. Particles are injected to replicate the natural flow expected of a complete disc. The distribution of the injected particles is derived from the isothermal calculations of \cite{batezeus}, interpolated to the pertinent protoplanet mass where necessary. This acts to replicate the expected gap in the vicinity of the protoplanet that would form were the entire disc modelled. The fact that the ZEUS calculations are vertically isothermal leads to a small mismatch in the vertical temperature structure at the boundaries for the radiation hydrodynamical calculations. These differences are small, particularly when compared with temperature changes that occur near the planet due to gas accretion; see Paper I for more details.

Ghost particles are employed along the boundaries of the disc section to provide the pressure forces that would come from the gas which is not modelled outside of the boundaries.

A boundary region above and below the disc, located at a height equivalent to an a optical depth of 1 ($\rm \tau \approx 1$) from infinity, allows the radiative transfer scheme to transport energy from the bulk of the disc to the boundary as though radiating to infinity. SPH particles within the boundary regions are evolved normally, except that they interact with the SPH particles in the bulk of the disc without their temperatures being affected.

\subsection{Equation of state}

We use an ideal gas equation of state for our radiation hydrodynamical calculations; $p=\rho T_{g} R_{g}/\mu$, where $R_{g}$ is the gas constant, $\rho$ is the density, $T_{g}$ is the gas temperature, and $\mu$ is the mean molecular mass.  The radiative transfer acts such that work and artificial viscosity (which includes both bulk and shear components) increase the thermal energy of the gas, and work done on the radiation field increases the radiative energy which can be transported via flux-limited diffusion. Energy transfer between the gas and radiation fields depends upon their relative temperatures, the gas density, and the gas opacity. For more details of the implementation see \cite{WhiBatMon2005} and \cite{WhiBat2006}.

\subsection{Opacity treatment}

We use interpolation through the opacity tables of \cite{pollack85} to provide the interstellar grain opacities (IGO) for solar metallicity molecular gas, whilst the gas opacities, which become important at high temperatures, are obtained from the tables of \cite{alexander75} (the IVa King model); for further details see \cite{WhiBat2006}. The agglomeration of interstellar grains into larger grains, or the sublimation of small grains, may lead to grain opacities that are lower than the IGO \citep{podolak2003}.  We mimic these effects by dividing the IGO values of \cite{pollack85} by factors 10, 100, or 1000 (details in Paper I).

\subsection{Protoplanets}

Our protoplanets are modelled by a gravitational potential, and a surface potential that yields an opposing force upon gas within one protoplanet radius of the protoplanet surface. The combination of the gravitational and surface forces takes the form of a modification to the usual gravitational force as
\begin{equation}
F_{r} = - \frac{GM_{p}}{r^{2}}\left(1 - \left(\frac{2R_{p}-r}{R_{p}}\right)^{4}\right)
\label{eq:surface}
\end{equation}
for $r <2~R_{\rm p}$ where $r$ is the radius from the centre of the protoplanet, $R_{\rm p}$ is the radius of the protoplanet, and $M_{\rm p}$ is the mass of the protoplanet. This equation yields zero net force between a particle and the protoplanet at the surface radius $R_{\rm p}$, whilst inside of the protoplanet's radius the force is outwards and increases rapidly with decreasing radius.

The majority of our calculations use protoplanets with radii equivalent to 1\% of their respective Hill radii (at 5.2 AU), but in one instance we model a Jupiter mass protoplanet with a radius equivalent to present day Jupiter ($\rm 7.15 \times 10^{9} cm$). A smooth start to the radiation hydrodynamical calculations is given by exponentially shrinking the protoplanet radii from an initial value of 0.01~$r_{\rm p}$ down to the desired radii during the first orbit of the protoplanet. The models settle to have steady temperature and density profiles after less than 2 orbits, with changes thereafter being very gradual (see Paper I, Figure 4).

\section{Results}
\label{results}

\begin{figure*}
\centering
\includegraphics[width=16cm]{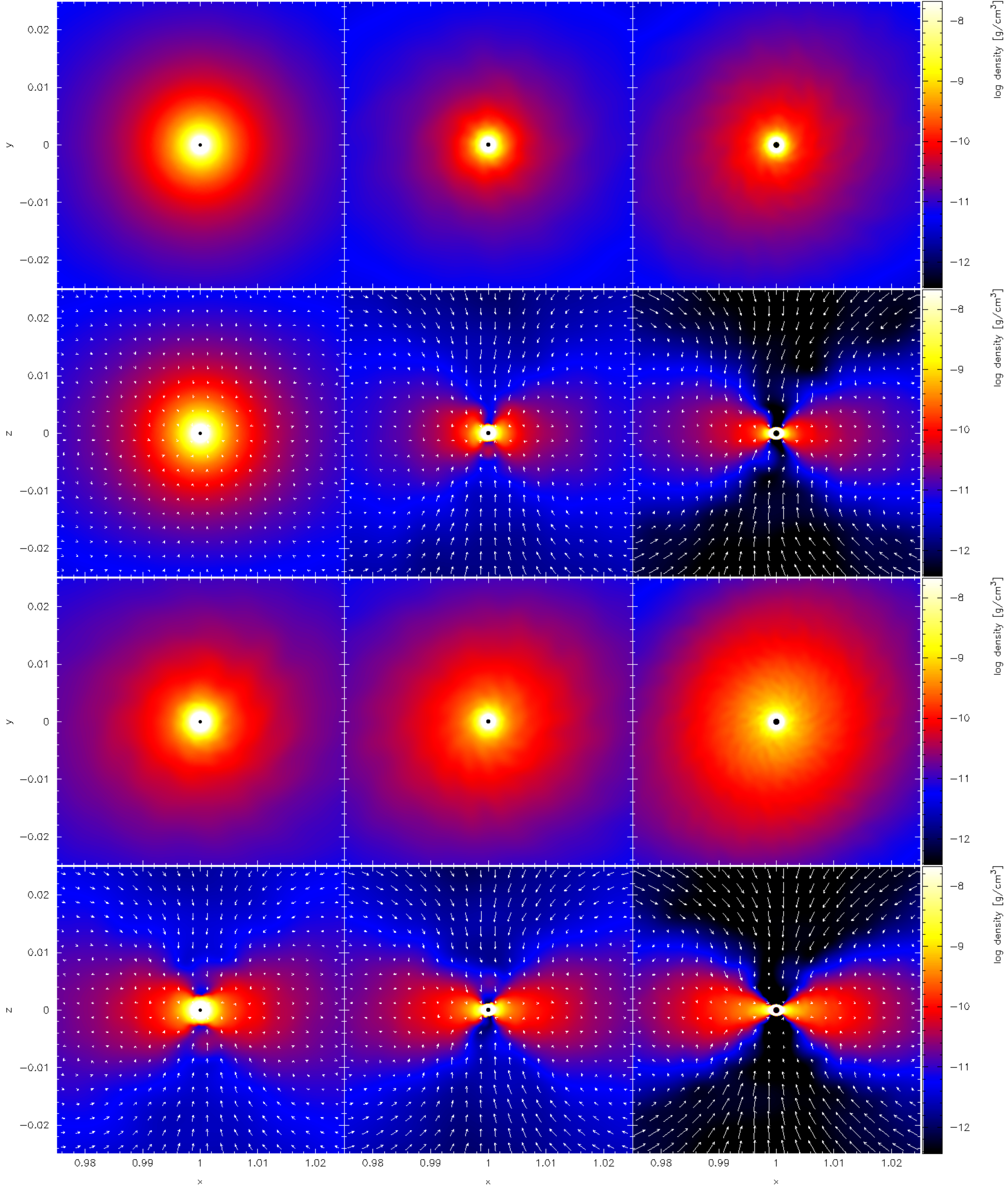}
\caption{Midplane and vertical density profiles of the gas distributions surrounding protoplanets with masses of 100, 166, and 333~$\rm M_{\earth}$ (left to right) and two different grain opacities. Top row: Density at the midplane in a nebula with interstellar grain opacity. Second row: Density in a vertical slice through the protoplanets, corresponding with row above. Third row: Density at the midplane in a nebula with 1\% interstellar grain opacity. Bottom row: Density in a vertical slice through the protoplanets, corresponding with row above. The spatial dimensions in units of the orbital radius (5.2 AU).}
\label{fig:discs}
\end{figure*}

We modelled protoplanets ranging in mass from 10-333~$\rm M_{\earth}$. Figure \ref{fig:discs} illustrates the density distribution about 100, 166, and 333~$\rm M_{\earth}$ protoplanets in both the $x-y$ and $x-z$ planes. We focus here on the immediate vicinity of the protoplanet. The reader interested in the structure on larger scales such as the spiral arms and gap in the circumstellar disc should refer to Paper I. The distributions about a 100~$\rm M_{\earth}$ protoplanet in a nebula with interstellar grain opacities (IGO) are shown in the top-left panels. The density distributions in the $x-y$ and $x-z$ slices are very similar; this case appears to be spherically symmetric, and the protoplanet is seen to possess no circumplanetary disc. However, the density distributions about a 100~$\rm M_{\earth}$ protoplanet in a 1\% IGO nebula, shown in the bottom-left panels of Figure \ref{fig:discs}, make evident the existence of a circumplanetary disc.

\begin{figure*}
\centering

\subfigure 
{
\setlength\fboxsep{0pt}
\setlength\fboxrule{0.0pt}
\fbox{\includegraphics[width=6.8cm]{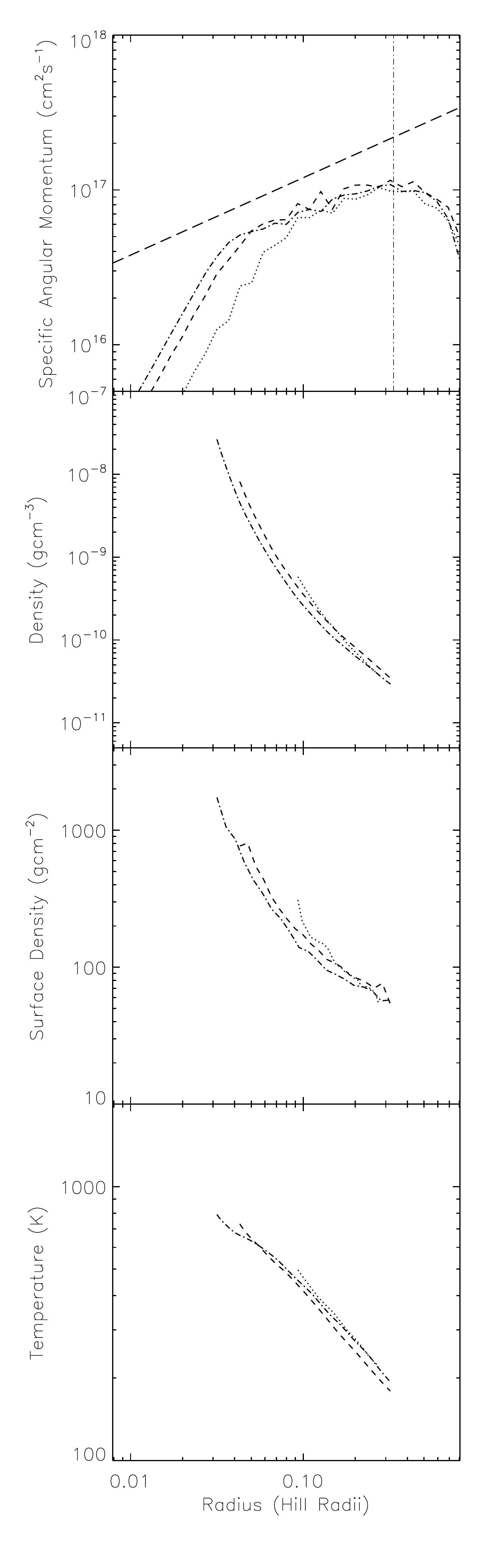}}
}
\nolinebreak
\hspace{-24.3mm}
\subfigure 
{
\setlength\fboxsep{0pt}
\setlength\fboxrule{0.0pt}
\fbox{\includegraphics[width=6.8cm]{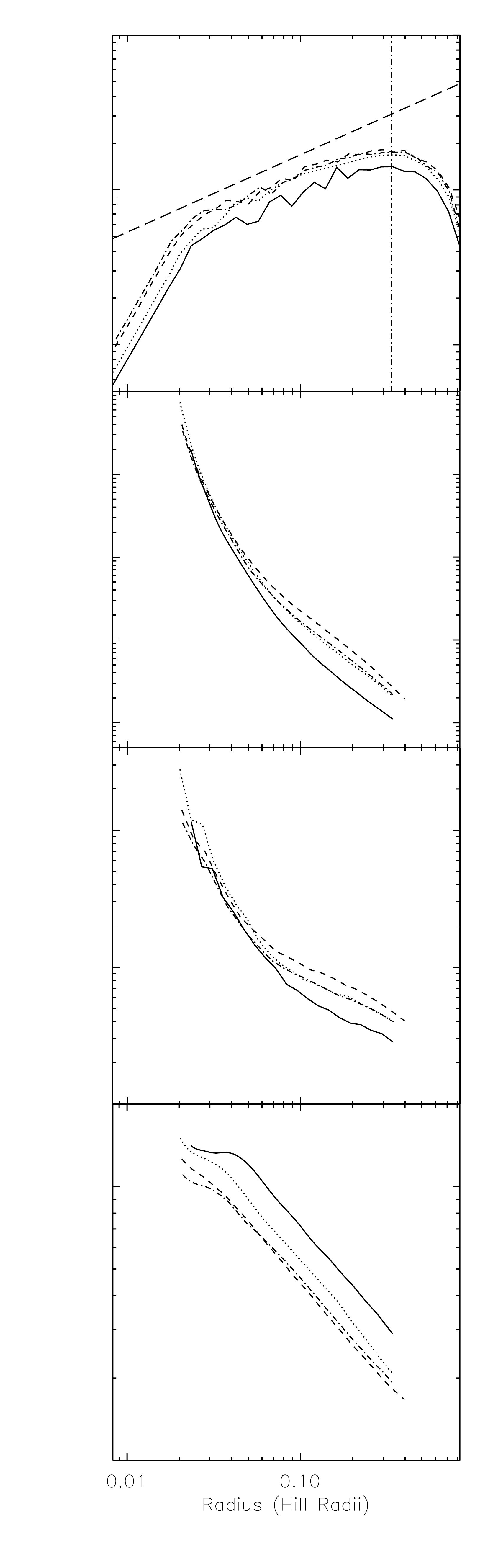}}
}
\nolinebreak
\hspace{-24.3mm}
\subfigure 
{
\setlength\fboxsep{0pt}
\setlength\fboxrule{0.0pt}
\fbox{\includegraphics[width=6.8cm]{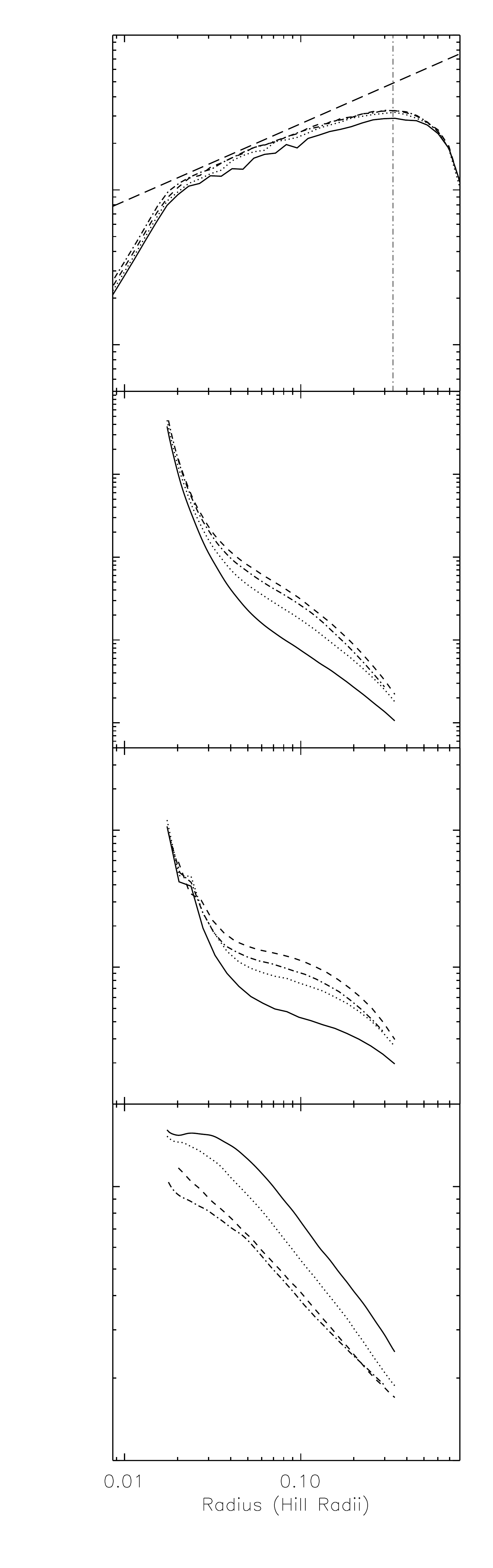}}
}
\vspace{-5mm}
\caption{Specific angular momentum (top row), midplane density (second row), surface density (third row), and midplane temperature (bottom row) profiles of the gas distributions surrounding protoplanets with masses of 100, 166, and 333~$\rm M_{\earth}$ (left to right) in nebulae of different opacities; IGO (solid line), 10\% IGO (dotted), 1\% IGO (dashed), and 0.1\% (dot-dashed). The straight dashed line in each panel in the top row indicates the expected specific angular momentum for material in Keplerian orbit about the protoplanet. The vertical dot-dash lines mark $R_{\rm Hill}/3$.}
\label{fig:discdens300}
\end{figure*}

To determine the size of the circumplanetary discs formed in our models we adopt a specific set of criteria. The radial edge of a circumplanetary disc is taken as the point of turnover in the specific angular momentum of the disc, calculated from the material within $\rm 5^{\circ}$ of the midplane. The top panels of Figure \ref{fig:discdens300} show the specific angular momentum distributions of gas around different protoplanets, in nebulae of different grain opacities. Furthermore, the absolute value of the angular momentum at the turnover point can be no more than a factor 2.2 different from the expected Keplerian specific angular momentum (denoted by straight dashed lines in Figure \ref{fig:discdens300}) at the given radius. The absolute difference in the specific angular momentum is due to pressure support which reduces the velocity that the gas must have to maintain a particular orbital radius.

\begin{figure}
\centering
\includegraphics[width=8cm]{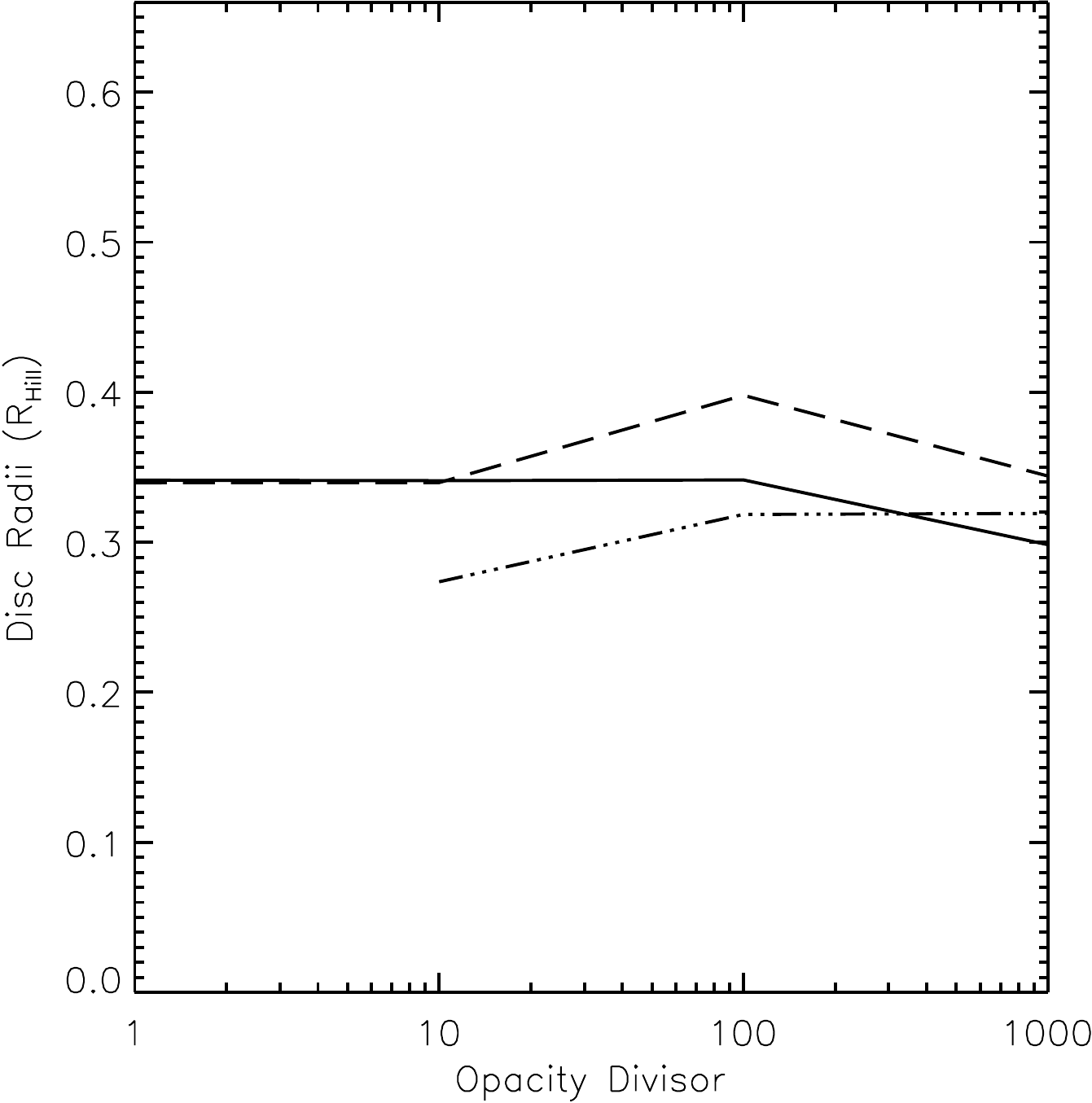}
\caption{The radial extent of circumplanetary discs around 100~$\rm M_{\earth}$ (dash-dots-dash line), 166~$\rm M_{\earth}$ (dashed line), and 333~$\rm M_{\earth}$ (solid line) protoplanets against varying grain opacity for the nebula material.}
\label{fig:discradiiopacity}
\end{figure}

Using the described criterion to determine whether a particular protoplanet has a disc, and the size of that disc, we found that only the 100, 166 and 333~$\rm M_{\earth}$ protoplanets possess circumplanetary discs. Figure \ref{fig:discradiiopacity} illustrates the circumplanetary disc radii obtained for these protoplanets. Protoplanets of lower mass exert an insufficient vertical gravity component to overcome the thermal pressure in their surrounding envelopes, thus they remain spherical. Just as was seen in Figure \ref{fig:discs}, the 100~$\rm M_{\earth}$ protoplanet in an interstellar grain opacity nebula is not found to have a circumplanetary disc. A simple analytic prediction of the approximate circumplanetary disc radii around accreting protoplanets was made by \cite{QuilTril1998}; their derivation is as follows. Assuming that the protoplanet is embedded in a gap, as is the case for the large masses described here, the mass inflow into the Hill sphere will enter via the inner and outer Lagrange points. An estimate of the specific angular momentum of the gas at these points can be made by assuming that the gas' transverse velocity component, $v_{\perp}$, comes purely of the Keplerian shear. In the rotating frame of the star-planet system
\begin{equation}
v_{\perp} = \sqrt{\frac{GM_{\star}}{r}} - \sqrt{\frac{GM_{\star}}{r_{\rm p}}},
\end{equation}
and the Lagrange points are approximately at
\begin{equation}
r = r_{\rm p} \pm R_{\rm Hill},
\end{equation}
where the Hill radius is
\begin{equation}
R_{\rm Hill} = r_{\rm p}\left(\frac{M_{\rm p}}{3M_{\star}}\right)^{1/3}.
\end{equation}
Thus,
\begin{equation}
v_{\perp} = \sqrt{\frac{GM_{\star}}{r_{\rm p}}}\left(\left(1+\frac{R_{\rm Hill}}{r_{\rm p}}\right)^{-\frac{1}{2}} -1\right).
\label{eq:expandable}
\end{equation}
Applying an expansion to the term with the form $(1 + x)^{-1/2}$ in equation \ref{eq:expandable} we obtain
\begin{equation}
v_{\perp} \simeq R_{\rm Hill}\sqrt{\frac{GM_{\star}}{r_{\rm p}^{3}}} = R_{\rm Hill}\Omega_{\rm p}.
\label{eq:velocity}
\end{equation}
Thus, an expression for the specific angular momentum of gas at the Lagrange points relative to the protoplanet is $j \simeq R_{\rm Hill}v_{\perp} \simeq R_{\rm Hill}^{2}\Omega_{\rm p}$. Assuming conservation of angular momentum, we estimate the centrifugal radius, $R_{\rm c}$, of the circumplanetary disc as
\begin{equation}
\frac{j^{2}}{R_{\rm c}^{2}} \approx \frac{GM_{\rm p}}{R_{\rm c}},
\label{eq:centrifugal}
\end{equation}
which can be solved to give
\begin{equation}
R_{\rm c} \approx R_{\rm Hill}/3.
\end{equation}

\noindent In all cases where circumplanetary discs are found, the disc radii do not differ greatly from this $R_{\rm Hill}/3$ estimate for a protoplanet embedded in a gap (Figure \ref{fig:discradiiopacity}). It is interesting to note that reducing the nebula grain opacity may enable disc formation, but that beyond that, changing the opacity appears to have only a small impact on the resulting circumplanetary disc radius.

In the second row of Figure \ref{fig:discdens300} we show the midplane density profiles of each circumplanetary disc. Excluding the inner most region where the density turns up due to gas piling upon the protoplanet, each disc density profile can be approximately described by a single power law. The exponent varies between -3/2 and -2 for different mass protoplanets and for nebulae of different grain opacities. The change of exponent with protoplanet mass and nebula opacity does not show an illustrative trend in either regard. The surface density distributions of the circumplanetary discs are shown in the third row of Figure \ref{fig:discdens300}, and follow a similar downward trend with radius, but with a shallower falloff ($\sim r^{-1/2}$ consistent with a constant $H/r$; see below).

\begin{figure}
\centering
\includegraphics[width=6.8cm]{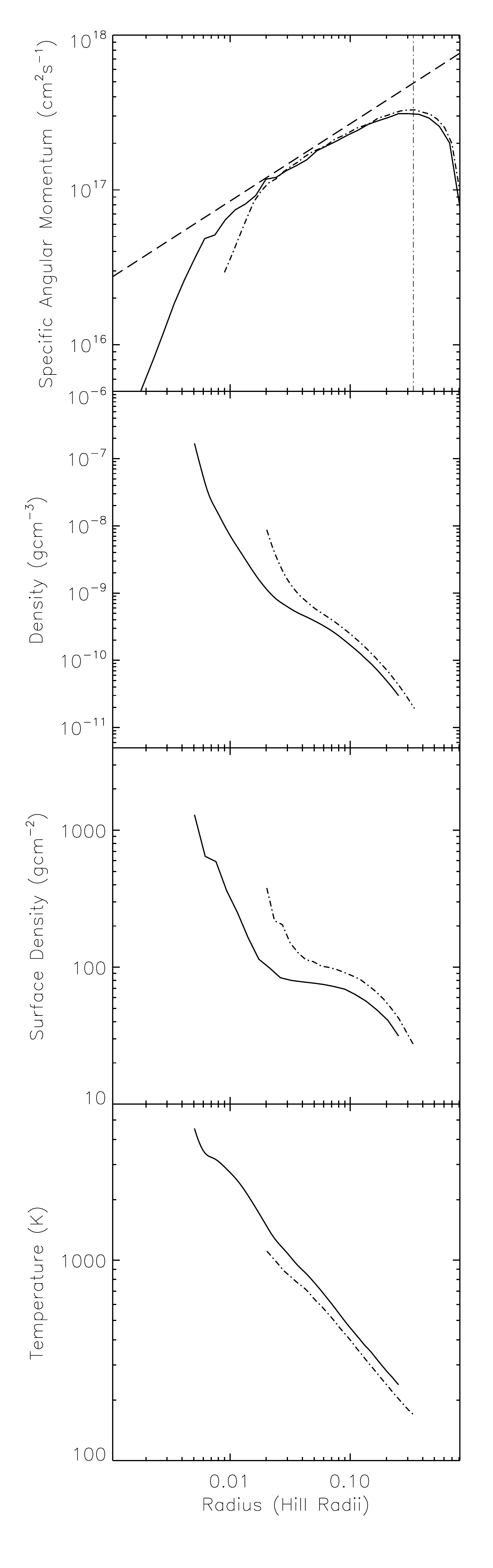}
\caption{The same quantities as plotted in Figure \ref{fig:discdens300}, but for our 333~$\rm M_{\earth}$ protoplanet with a surface at a Jupiter radius instead of 1\% $R_{\rm Hill}$. The dot-dash line gives the results for the 1\% $R_{\rm Hill}$ case at an equivalent time for comparison; this is at an earlier time than the 1\% $R_{\rm Hill}$ case in Figure \ref{fig:discdens300} as the small core calculations are slower, and thus less evolved. The nebula has 1\% interstellar grain opacities.}
\label{fig:midplane}
\end{figure}

The bottom row in Figure \ref{fig:discdens300} shows the temperature profiles calculated by azimuthally averaging the temperature of gas at the midplane of the circumplanetary discs. The temperatures are relatively high, ranging from 170K to 1600K. The lowest temperatures are achieved at reduced grain opacities. The fall off in temperature with radius for all opacities and protoplanet masses is approximately described by a power-law, $r^{-7/10}$. For our Jupiter radius protoplanet calculation the disc's inner edge is nearer to the protoplanet's centre, and thus within a steeper potential leading to a correspondingly higher temperature, up to $\rm \sim 4500K$. The outer edge has a temperature of 240K, only slightly larger than obtained for the same calculation performed with a 1\% $R_{\rm Hill}$ surface radius (Figure \ref{fig:midplane}). Thus the chosen size of the protoplanet has only a small effect on conditions further out in the circumplanetary disc, and does not change our main conclusions.

\begin{figure}
\centering
\includegraphics[width=8cm]{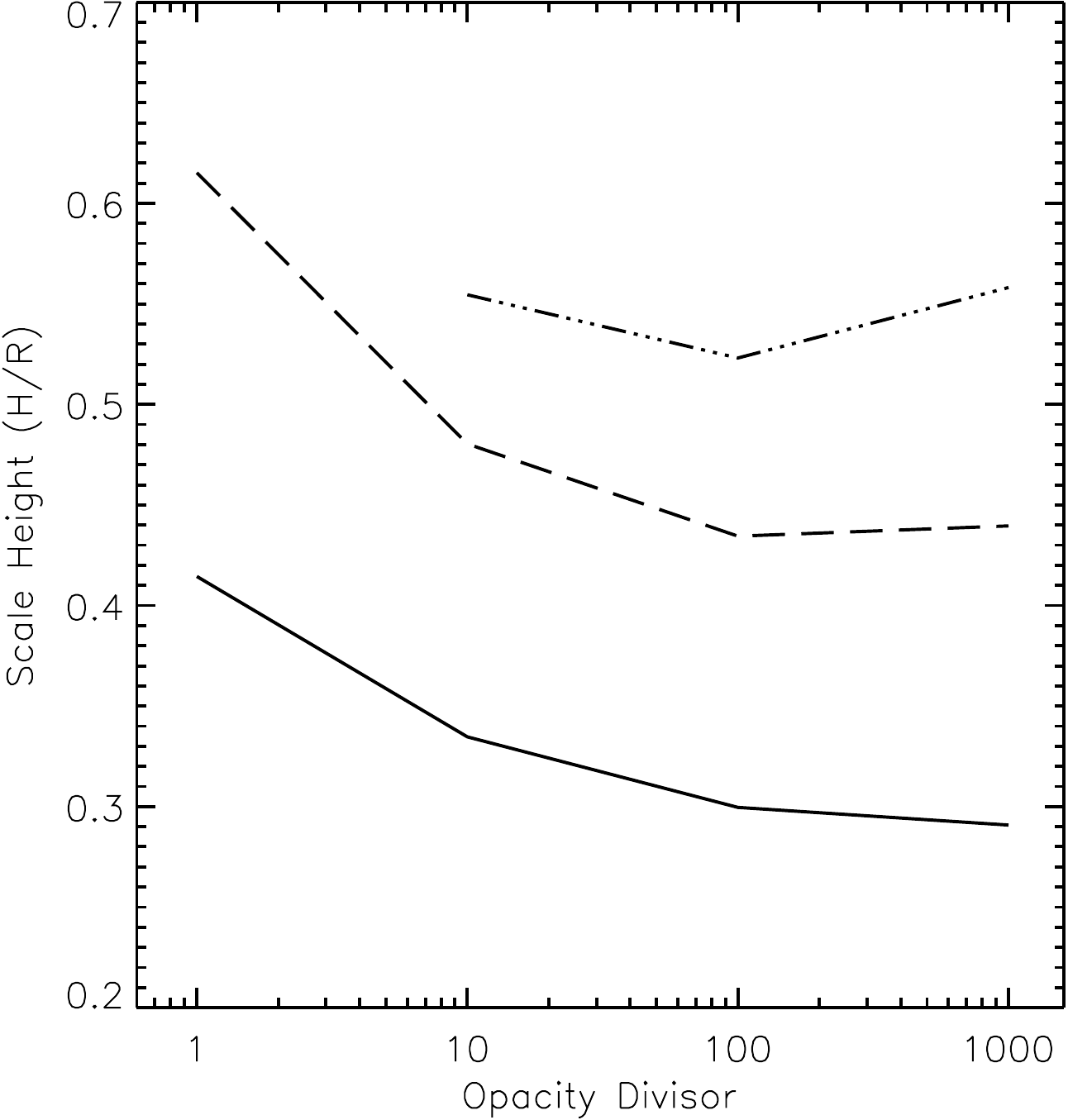}
\caption{Scaleheights of circumplanetary discs around 100~$\rm M_{\earth}$ (dash-dots-dash line), 166~$\rm M_{\earth}$ (dashed line), and 333~$\rm M_{\earth}$ (solid line) protoplanets against varying grain opacity for the nebula material.}
\label{fig:scaleheights}
\end{figure}

All of the circumplanetary discs that are produced in our models are thick. The scaleheights $(H/r)$ are shown in Figure \ref{fig:scaleheights} and can be seen to range between $\sim 0.3 - 0.6$. The scaleheights are constant with radius, showing no kind of flaring. Scaleheight tends to decrease with decreasing grain opacity due to the greater radiative cooling possible at lower opacities. This leads to cooler discs (as can be seen in Figure \ref{fig:discdens300}) and therefore more flattened circumplanetary discs (as can be seen in Figure \ref{fig:discs}). Only the lowest grain opacity does not follow this trend as the gas and radiation can decouple at such low opacities, reducing the effectiveness of radiative cooling. The scaleheights also tend to decrease with increasing protoplanet mass due to the larger vertical gravity component; once again, this effect is visible in the $x-z$ slices in Figure \ref{fig:discs}. The scaleheight for the disc about our Jupiter radius protoplanet is broadly in line with the other 333~$\rm M_{\earth}$ protoplanet scaleheights, however it is slightly thinner than its 1\% $R_{\rm Hill}$ surface radius counterpart; $H/r \simeq$ 0.3 and $H/r \simeq$ 0.4 respectively.

\begin{figure}
\centering
\includegraphics[width=8cm]{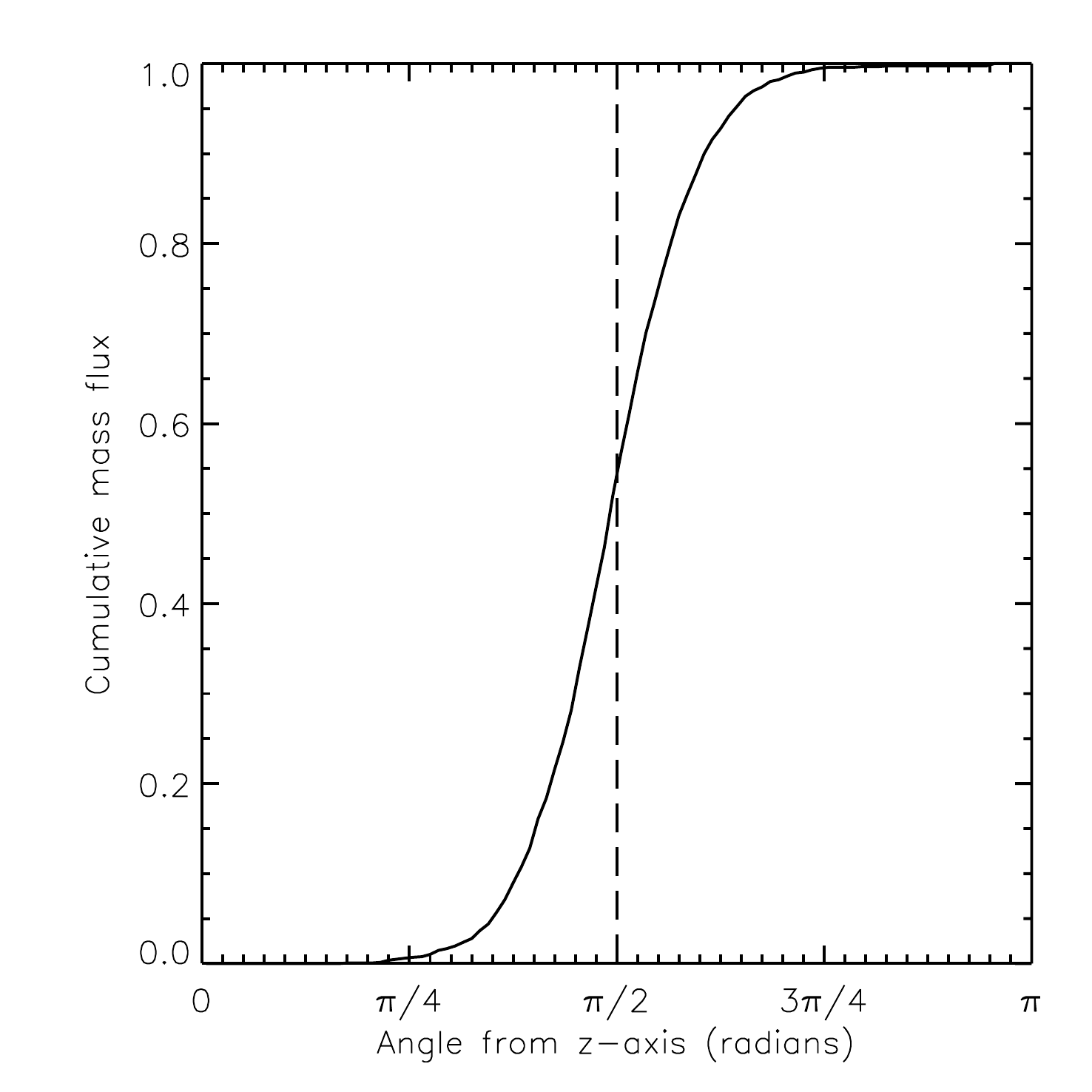}
\caption{We investigate the latitudinal dependence of the mass inflow rate through the Hill sphere on to a 333~$\rm M_{\earth}$ protoplanet embedded in a nebula with interstellar grain opacity.  We plot the normalised cumulative distribution of mass flux against latitude (the disc midplane is at $\pi/2$).  The vast majority of the mass flows into the Hill sphere near the equator.}
\label{fig:massflux}
\end{figure}

The $x-z$ plane density plots in Figure \ref{fig:discs} also include velocity vectors, to indicate the relative velocity of gas flow in the vicinity of the protoplanet. The gas velocities down on to the disc and protoplanet from high latitudes are considerably greater than the velocities through the disc. This velocity structure, with fast infall of gas from high latitudes, reflects what was found by \cite{KlahrKley2006}. However, the gas above and below the midplane is very diffuse, thus the mass flux in this direction contributes a very small fraction to the growing protoplanet. This can most clearly be seen in Figure \ref{fig:massflux} which illustrates that a majority of the mass enters the Hill sphere of the protoplanet at around $\rm 90^{\circ}$ from the $z-$axis (i.e. near the $x-y$ plane), feeding in to the circumplanetary disc.

\section{Discussion}
\label{discussion}

The properties of circumplanetary discs around protoplanets are important in understanding the formation and capture of the satellite systems we see around giant planets in our solar system.

\subsection{Comparison with previous work}

A massive protoplanet that establishes a well defined gap in its parent circumstellar disc is expected to accrete material that flows into that gap via its Lagrange points. This gas in turn is expected to have low velocity relative to the protoplanet upon entering the gap due to its similar orbital radius with respect to the star. Using these assumptions \cite{QuilTril1998} estimated a centrifugal radius of $\sim R_{\rm Hill}/3$; at this radius gas is expected to enter into orbit of the protoplanet. Gas must then evolve towards the protoplanet by angular momentum transport, forming a circumplanetary disc of material as it slowly infalls. The radial extents of the circumplanetary discs in our models are in good agreement with this estimate, and markedly larger than those assumed in the models of \cite{CanWar2002}. \citeauthor{QuilTril1998} also give an estimate for the expected disc radii about protoplanets that have not evacuated a disc gap. In these instances the gas flowing into the vicinity of a protoplanet has lower specific angular momentum, and the resulting circumplanetary disc is considerably smaller. However, we find no clear evidence of circumplanetary discs around protoplanets of less than 100~$\rm M_{\earth}$, instead we find near spherical envelopes.

\cite*{DAnHenKle2003} performed two-dimensional hydrodynamical calculations to investigate circumplanetary discs around protoplanets, including viscous heating and simplified radiative cooling. For a Jupiter mass protoplanet, they also find circumplanetary discs with outer edge radii (as defined by a turnover in specific angular momentum) of around $R_{\rm Hill}/3$, inline with \cite{QuilTril1998} and our own results. The initial properties of our circumstellar disc (in which our planets are embedded) are similar to those of \cite{DAnHenKle2003}, with an identical surface density falloff, and similar initial temperatures and densities at the planet's orbital radius. The circumplanetary disc properties that develop in our calculations are also similar. For our Jupiter mass protoplanet, the circumstellar disc at standard opacity reaches a maximum temperature of 1600K at $\approx 0.02 R_{\rm Hill}$, whilst \citeauthor{DAnHenKle2003} find a maximum temperature of 1500K at the same radius. The surface density profiles in both sets of models are also similar. 

\cite{Machida2008} found that the circumplanetary discs formed in his calculations possessed a peak in surface density around 30 protoplanet radii from the protoplanet's centre. Figure \ref{fig:midplane} shows the surface density and azimuthally averaged midplane density against increasing radius for our calculation of a Jupiter mass protoplanet with a Jupiter radius. The midplane density, and likewise the surface density, always decrease with increasing radius from the protoplanet out to the Hill radius with no sign of a ring-like density enhancement. This difference between Machida's results and our own is most likely a consequence of our differing treatments of the gas nearest the protoplanet. Machida uses a sink cell to remove gas at a radius approximately equal to a Jupiter radius, creating an evacuated central region that leads to an inward pressure gradient near the protoplanet. Our more realistic approach is to model a surface upon which the gas piles up, increasing in density and temperature, and exerting an outward force on gas external to it.

Our three-dimensional models show no evidence of strong shocks within the circumplanetary discs, in agreement with both \cite{batezeus} and \cite{Machida2008}, and in contrast to previous two-dimensional models. This again suggests that the strong shocks seen in two-dimensional models are produced by a lack of vertical spread in the colliding gas streams. A lack of strong shocks may be more conducive to the agglomeration of solids as the disc may be more quiescent. Shock driven migration of material through the discs may also be reduced, leading to longer lived discs. However, in contradiction to these previous three-dimensional calculations, we find very much thicker circumplanetary discs. This is caused by the trapping of heat in the optically thick circumplanetary disc; unable to radiate away the heat, the disc puffs up. A hotter disc may lead to faster accretion, opposing any enhanced lifetime gained from the weakened shocks.


\subsection{Implications for satellite formation}

The temperatures found in our circumplanetary discs are quite high, particularly when considering the large fraction of ice found in some of the Galilean moons. The inner most Galilean satellite, Io, is mostly composed of silicates and metals, Europa, mostly silicates with $\rm \sim 10\%$ ice, whilst Ganymede and the outermost Galilean satellite, Callisto, both include some 50\% ice \citep{SchSpoRey1986}. The usual interpretation of the increasing ice fraction with orbital radius is of a radially decreasing temperature within the progenitor circumplanetary disc \citep{LunSte1982}. This temperature profile must reach a low enough temperature around Ganymede's orbit for large amounts of water ice to form and be incorporated in to the forming satellites \citep{CanWar2002}. However, only at the outermost reaches of our circumplanetary discs at reduced opacities does the temperature approach the $\rm \sim 160K$ expected for water condensation (at 5AU around a solar type star, \citealt{Stevenson1989}). This is at a radius some 17 times the present orbital radius of Ganymede. This may point towards orbital migration, though we have shown that there is a dearth of material at very large radii within a circumplanetary disc, which makes satellite formation more challenging further from the protoplanet. It is also possible that satellite formation occurs after significant gas accretion from the circumstellar disc has ceased, and the residual disc has cooled. Alternatively, perhaps ice must be delivered from the surrounding solar nebula in sufficiently large bodies such that it does not sublimate prior to satellite growth.

It could also be that the moons were captured. However, the capture of the Galilean satellites is not supported by their low inclinations and eccentricities, which are suggestive of formation within a circumplanetary disc. However some of Jupiter's satellites which have both low inclinations and eccentricities do appear to have formed in locations other than their present orbits, for example Amalthea \citep{Cooper2006}. Amalthea orbits inside of Io, and yet has a density lower than water \citep{Anderson2005}, suggesting it formed elsewhere. However its low inclination and eccentricity can be explained through its interactions with Io \citep{Burns2004}, and needn't have been properties of its formation, as is supposed for the Galilean satellites. For a recent review of the issues surrounding satellite formation in circumplanetary discs, see \cite{Estrada2008}.

It should be noted that we do not find circumplanetary discs around protoplanets with masses similar to Uranus and Neptune. This may not be a problem for Neptune, since Triton, Neptune's primary satellite has a retrograde orbit. However, Uranus' satellite system exhibits properties that are suggestive of a disc origin. Again, this might be resolved by considering the eventual cooling of such a protoplanet's rotating envelope following the dispersal of the encompassing circumstellar disc. As the envelope cools it may well flatten into a disc suitable for satellite growth.

\section{Conclusions}

We have investigated the effects of grain opacity and protoplanetary mass upon the properties of circumplanetary discs. We performed calculations for protoplanets ranging in mass from 10-333 $\rm M_{\earth}$, with grain opacities from 0.1\% to full interstellar values. Our protoplanets have radii of 1\% their respective $R_{\rm Hill}$, excepting one comparison calculation of a Jupiter radius.

As was stated in \cite{AylBate2009} we find that 100~$\rm M_{\earth}$ protoplanets at reduced opacities (10, 1, and 0.1\% interstellar grain opacities) possess circumplanetary discs, as do all the higher protoplanet masses that we model at all opacities. The radial extent of all these circumplanetary discs are in good agreement with the analytical prediction for discs formed in gaps evacuated by high mass protoplanets of one third of the Hill radius \citep{QuilTril1998}. Such large discs might contribute significant flux to direct imaging detections of young exoplanets by reflecting light from the central star. Indeed, it has been proposed that the recent imaging of the planet orbiting Formalhaut \citep{Kalas2008} may have included significant flux from a circumplanetary disc.

The discs produced are all thick in comparison with their parent circumstellar discs, and their scaleheights are found to vary with opacity. They become thicker at higher grain opacities, as the thermal pressure supports them more against the vertical gravitational pull of the protoplanet. This gravitational influence in turn leads to a scaleheight trend with protoplanet mass, the more massive protoplanets possessing more flattened discs.

We have found that the temperatures achieved in the circumplanetary discs are high when compared with the condensation temperature of ice. Only at the outer edges of our discs, and then only in reduced grain opacity cases, does the temperature fall low enough for ice formation. This poses a problem vis-\`{a}-vis the formation of icy satellites around gas giant planets.

Finally, in contrast to the three-dimensional models of circumplanetary discs of \citet{Machida2008}, we fail to find a ring-like density enhancement within the circumplanetary disc. Rather, we find that the density decreases monotonically with radius from the protoplanet.

\section*{Acknowledgments}

We are grateful to the anonymous referee for their useful suggestions. BAA and MRB would also like to thank Jenny Patience and David Rundle for many useful conversations. The calculations reported here were performed using the University of Exeter's SGI Altix ICE 8200 supercomputer. MRB is grateful for the support of a Philip Leverhulme Prize and a EURYI Award. This work, conducted as part of the award ``The formation of stars and planets: Radiation hydrodynamical and magnetohydrodynamical simulations"  made under the European Heads of Research Councils and European Science Foundation EURYI (European Young Investigator) Awards scheme, was supported by funds from the Participating Organisations of EURYI and the EC Sixth Framework Programme. Many visualisations were produced using SPLASH \citep{splash}, a visualisation tool for SPH that is publicly available at http://www.astro.ex.ac.uk/people/dprice/splash.

\bibliography{paper.bib}

\end{document}